# TOWARDS A NEW FRAMEWORK LINKING KNOWLEDGE MANAGEMENT SYSTEMS AND ORGANIZATIONAL AGILITY: AN EMPIRICAL STUDY

Mohamed Amine Marhraoui [1] and Abdellah El Manouar [2]

ENSIAS Engineering School, Mohammed V University in Rabat

## ABSTRACT

*The amount of data has exploded over the last ten years. Data is captured and shared from personal devices, transactional operations, sensors, social media and other sources. Firms should, thus, be able to explore the new opportunities and rapidly seize them by developing the corresponding capabilities. In our work, we focus on two emerging dynamic capabilities: Absorptive capacity and organizational agility. We propose a new theoretical Framework based on the previous literature linking the use of knowledge management systems and firm's organizational agility by highlighting the mediating role of firm's absorptive capacity. In addition, we carried out an empirical study based on a survey to support and validate the proposed Framework. The main findings of this study are presented.*

## KEYWORDS

## 1. INTRODUCTION

In the recent years, the global economy has evolved to an economy of knowledge. The immaterial capital of a firm is becoming crucial for its development. Indeed, the post industrial society is characterized by an increasing importance of knowledge rather than infrastructure or capital [1].

The 21st century is marked by the important place of knowledge workers in the society instead of manual workers in manufacturing during the 20th century. These manual workers are considered as a cost for the company. However, knowledge workers are an essential capital asset for the growth of the firm [2], as knowledge is becoming increasingly important for enhancing firm's performance ([3], [4]).

The transition from economies based on manufacturing to others based on services has increased the production of knowledge. [5].

Indeed, all OECD (The Organisation for Economic Co-operation and Development) countries have moved from models based on labour, raw materials and physical capital to economies of knowledge and intangible capital [6].

The knowledge economy is closely linked to globalization, to the use of advanced technologies, to the development of international trade and financial services (capital markets, global payments, etc...) [7].

Keith Smith (2002) suggests that the growing importance of knowledge in an industry is essentially related to the distributed knowledge base across people, organizations and technologies rather than the internal firm's knowledge [8].

Philip Brown (2008) confirms that knowledge quality is decisive for the competitiveness of countries. Companies gain an advantage based on the highly skilled workers in developing countries like China/India and by standardizing knowledge work. This confers them with high quality knowledge for fewer prices [9].

Following researches of the World Bank Institute (2007) [10], the knowledge economy relies on four pillars:

1) Institutional support for the use of new and existing knowledge by offering an adequate environment which facilitates entrepreneurship, foreign direct investments and international trade.
2) Skilled human resources able to diffuse and exploit new knowledge.
3) Effective infrastructure and information technologies / networks to store and explore knowledge.
4) Academic partnerships with R&D centres and innovation clusters.

In this article, we will focus on the third pillar and especially on knowledge management systems (KMS). We distinguish collaborative KMS and decision-oriented KMS.

Also, as the importance of knowledge is ever more important in a fast changing environment, we study two dynamic capabilities: absorptive capacity and organizational agility.

The structure of this work is organized as follow. Section Two describes a literature review presenting knowledge, knowledge management and the related systems. Also, we provide a review of the two dynamic capabilities: absorptive capacity and organizational agility. Section three tackles the Framework proposal. Section four is devoted to the empirical study. We will analyse our survey results and present the main findings in relation to the impact of knowledge management systems on organizational agility and the intermediary role of absorptive capacity.

Finally, we provide a conclusion of the article and highlight our research perspectives.

## 2. LITERATURE REVIEW

### 2.1. KNOWLEDGE MANAGEMENT SYSTEMS

#### 2.1.1. DATA, INFORMATION AND KNOWLEDGE:

Knowledge is related to the specific ideas or understandings that a firm has created and used to achieve its goals [11].

Knowledge can be either explicit when it is rational, sequential or digital; or implicit/tacit when it is related to experience, practice and context [12].

There are important characteristics of knowledge, as a resource, which allow firms to create value: transferability, aggregation potential and appropriability especially of tacit knowledge [13].

Knowledge is different from data and information. Indeed, data is related to raw numbers, characters and words. It is objective and informs about facts [14].

Data can be either structured or unstructured when it doesn't fit an established schema [15].

The unstructured data is stocked in non relational databases called NoSQL databases [16].

This large amount of structured and unstructured data constitutes what is called "Big Data". This latter is generated from different sources: customer's clicks, social media contents, commercial transactions, sensor objects… in addition to traditional static sources [17].

However, information is data when interpreted and processed giving it meaning.

In addition, knowledge is information when personalized in order to allow effective action [14].

Based on these differences, a hierarchy is established between data, information, knowledge and Wisdom. This latter is the capacity to choose the right knowledge in order to increase effectiveness and to respect values [18].

### 2.1.2. KNOWLEDGE MANAGEMENT:

Knowledge management is the process of identifying, creating and developing knowledge in the company in order to gain competitive advantage [19].

Knowledge management can be performed in different levels of the firm, particularly the individual level, team level and organizational level. [20]

From a resource based view of the firm, knowledge management aims to provide the company with valuable resources and difficult to imitate knowledge [24].

Thus, knowledge management can provide the company with the knowledge required to support the overall business strategy by assessing the current state of firm's knowledge and filling knowledge gaps [22].

### 2.1.3. KNOWLEDGE MANAGEMENT SYSTEMS:

Knowledge management systems are systematic mechanisms to managing knowledge. They allow the company to select and assimilate knowledge embodied in business processes [23].

Knowledge management systems include the IT infrastructure which enables the firm to create new knowledge, to store it, to share and diffuse it, and to apply it for effective actions [19].

Knowledge management systems are either centralized or peer to peer. On one hand, centralized systems are based on multi-layer architecture in order to access, to integrate and to structure knowledge from different sources (operational databases, data warehouses, documents repositories …). On the other hand, peer-to-peer systems provide the same services as the centralized ones, but are based on client/server architecture. The client side is related to personal knowledge and the server side provides centralized data and knowledge sources (CRM, ERP systems…) [24].

Based on the literature, we distinguish two main knowledge management systems (KMS) categories, namely collaborative knowledge management systems (CKMS) and decision-oriented knowledge management systems (DKMS) [25].

The first type is related to the capture and sharing of Knowledge. It includes: content management systems (CMS), document management systems (DMS), intranet/extranet portals, groupware and workflow tools [26].

The second type consists of the systems allowing decision-making based on knowledge discovery and relevancy. It is related to decision support systems based on data warehouses and data mining techniques. Also, artificial intelligence and machine learning enable firms to automatically learn to recognize complex patterns and make intelligent decisions [27]. This second type contains likewise advanced analytics based on big datasets [28].

## 2.2. ABSORPTIVE CAPACITY

The absorptive capacity of the firm is the ability to recognize valuable information and to exploit external knowledge [29]. Zahra and George [30] distinguish fours dimensions of absorptive capacity: Acquisition of relevant knowledge from external sources; assimilation of this knowledge by analyzing and understanding; transformation of knowledge converted in order to be combined with internal sources for new insights; and exploitation of the new knowledge within the company.

Acquisition and assimilation of external knowledge constitutes the potential absorptive capacity of the firm, and transformation/exploitation inside the company is named realized absorptive capacity [30].

## 2.3. ORGANIZATIONAL AGILITY

In 1992, Robert Nagel and Rick Dove wrote a report so as to boost the American economy. Published by the IACCOCA Institute, the aim of the report was to establish a strategy for the next 15 years allowing the US industry to regain its competitiveness and build a new model in place of mass production.

Organizational agility is the ability to detect opportunities of innovation, and to seize the opportunities of the market by preparing the required assets and knowledge in a rapidly manner [31]. It is the capacity of a company to adapt itself to the changes, often unforeseen, in its environment, and to exploit changes as opportunities of development and growth through fast and innovative answers [32]. Organizational agility helps the company to deal witch rapidly changing environment. This latter is characterized by volatility, uncertainty, complexity, and ambiguity (VUCA) [33].Volatility means the pace, speed and volume of change. Uncertainty deals with difficulties to predict events [34]. Complexity is related to the chaos embracing the company and ambiguity is associated with changing conditions and contexts [35].

Two components constitute the organizational agility of the firm: sensing and responding capabilities. The sensing capability allows the company to predict customer expectations and trends, to identify technology advancements and to deal with political and regulatory changes [36] .The responding capability is the company's ability to have different options to rapidly act by deploying existing resources or building new ones [37].

# 3. FRAMEWORK PROPOSAL

## 3.1. FRAMEWORK DESCRIPTION

Based on the literature, studies have focused on the impact of knowledge management on enhancing firm's organizational agility [38, 39].

Also, previous articles present the Information technology infrastructure as an enabler of firm's agility by enhancing its    ability to sense opportunities and to respond adequately.

Thus, we aim through our proposed framework to study the unexplored area of the relationship between knowledge management systems and organizational agility through absorptive capacity playing a mediating role (Figure. 1).

The proposed framework stipulates that firms investing in knowledge management systems (collaborative KMS and decision-oriented KMS) can improve their absorptive capacity. This latter enhances organizational agility of the firm.

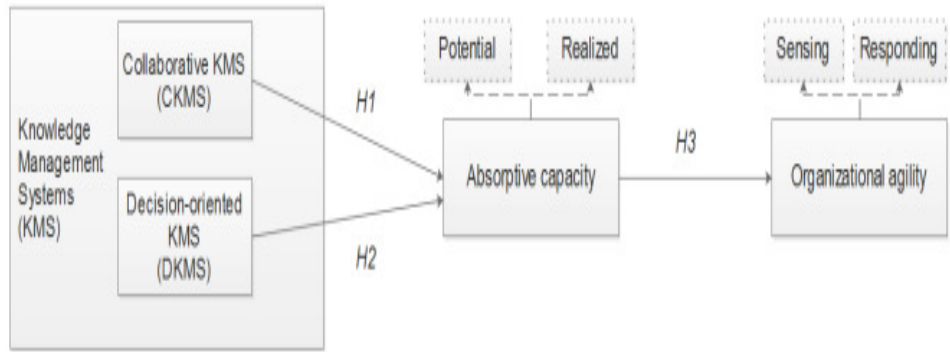

Figure. 1:  The proposed framework

## 3.2. HYPOTHESIS DESCRIPTION

The table below presents the main hypothesis of our proposed framework (Table 1.):

Table 1. Hypothesis of our proposed model.

| Hypothesis | Description |
|---|---|
| H1 | Firms investing in collaborative knowledge management systems (CKMS) improve their absorptive capacity |
| H2 | Firms investing in decision-oriented knowledge management systems (DKMS) improve their absorptive capacity |
| H3 | The more a firm has better absorptive capacity the more it is agile |

H1: Firms investing in collaborative knowledge management systems (CKMS) improve their absorptive capacity.

Collaborative knowledge management systems drive the assimilation of new knowledge in order to be embedded in organization processes and to be used. Assimilation is facilitated by knowledge storage tools (documents, diagrams ...), by converting into understandable knowledge format using intelligent agents (e-mail agents, scheduling agents), and by organizing knowledge into different ontologies [40].

In addition, collaborative knowledge management systems allow the transformation of knowledge within the same specialization or across different domains [41].
Knowledge is transferred through tools like discussion forums, electronic bulletin boards, computer networks, and corporate directories in order to allow locating persons or business units where to look for desired knowledge inside and outside the company [42].

H2: Firms investing in decision-oriented knowledge management systems (DKMS) improve their absorptive capacity.

Decision-oriented knowledge management systems enhance the potential (PACAP) and realized (RACAP) absorptive capacities of the firm [43].

Indeed, Data mining tools help to organize data by analysing dependencies, grouping into classes, comparing and summarizing data, detecting anomalies and visualizing in order to select relevant knowledge [44].

Also, business intelligence systems allow giving insights for decision makers based on automated reports for analysis and the integration of the precedent validated analysis outputs [45].

H3: The more a firm has better absorptive capacity the more it is agile.

A firm with developed absorptive capacity acquire in a timely manner relevant knowledge which can be used to sense opportunities and market changes [46].

As a VUCA environment can be marked by uncertainty and lack of information, absorptive capacity allows the firm to acquire relevant knowledge which can be useful to have insight about the current situation and the future [47].

Also, the absorptive capacity of the firm enhances its response ability. Indeed, the acquired external knowledge associated with an effective diffusion inside the company allows it to adapt continuously to environment uncertainty and turbulence [48].

This latter is related to the new introduced products, new customers and changing market conditions [49].

## 4. EMPIRICAL STUDY

### 4.1 METHODOLOGY AND DATA COLLECTION

We conducted a survey across different economy sectors (private and public) and size organizations (SMEs and large businesses) through a questionnaire which tackles the use of knowledge management systems and firm's capabilities.

The questionnaire was directly sent by e-mail to an initial database containing about one thousand of corporate contacts in different positions which constitutes the sampling pool of our survey.

In order to get reliable responses, the contact databases were provided by two training institutions which collaborate with executives, managers and non managers.

131 valid responses were received and used for the analysis which represents a response rate of about 13%.

Respondents are mainly located in the French-speaking area, especially: Morocco, France, Tunisia, Algeria, Senegal, Ivory Coast, and Togo…

The sample characteristics are detailed in Table 2.

Table 2. Sample Characteristics (N=131).

| Measure | Item | Percentage |
|---------|------|-----------|
| Gender | Man | 64.88% |
| | Women | 35.12% |
| Age | 18-29 | 24.43% |
| | 30-44 | 47.33% |
| | 45-59 | 27.48% |
| | 60-74 | 0.76% |
| Position | CEO | 6.11% |
| | Board Member | 7.63% |
| | Manager | 45.04% |
| | Non manager | 41.22% |
| Size | < 10 | 19.85% |
| | Between 10 and 100 | 25.95% |
| | Between 100 and 5000 | 25.95% |
| | +5000 | 28.25% |
| Sector | IT industry | 19.08% |
| | Finance services | 9.16% |
| | Consulting | 11.45% |
| | Construction | 3.05% |
| | Public administration | 3.82% |
| | Energy | 7.63% |
| | Others | 45.04% |
| Country | Morocco | 24.43% |
| | France | 52.67% |
| | Tunisia | 2.29% |
| | Algeria | 1.53% |
| | Ivory coast | 2.29% |
| | Germany | 2.29% |
| | Others | 14.50% |

Table 3. Below presents the constructs of our model, the items used and the corresponding literature.

Table 3. The constructs and items of our model

| Construct | Item | Literature |
|-----------|------|------------|
| CKMS | KMS1: use of groupware and workflow tools. | (Dave and Koskela, 2009) [50]; (A.Serenko et al., 2016) [51] |
| | KMS2: use of Intranet/extranet. | |
| | KMS3: use of DMS (Document Management System). | |
| | KMS4: use of CMS (Content Management System). | |
| DKMS | KMS5: decision support systems based on data warehouse. | ( C. Fredriksson, 2015) [52]; ( Greco et al., 2013)[23] |
| | KMS6: big data analytics initiatives. | |
| | KMS7: manage knowledge using AI (Artificial Intelligence). | |
| Organizational agility | OA1: R&D initiatives. | (Sambamurthy et al., 2003) [31]; (Covey et al., 2006) [36]; (Markides, 2006) [53]; (Brown and Eisenhardt, 1995).[54] |
| | OA2: market intelligence. | |
| | OA3: optimised production cycles. | |
| | OA4: new conception methods (Design thinking). | |
| | OA5: flexible organization. | |
| | OA6: exploit technological advancements. | |
| | OA7: involve suppliers in services. | |
| | OA8: integrate reconfigurable resources. | |
| Absorptive capacity | AC1: identify external knowledge. | (Cohen and Levinthal,1990) [29]; ( Zahra and George, 2002)[30] |
| | AC2: interpret new knowledge | |
| | AC3: combine new and existing knowledge. | |
| | AC4: apply new knowledge. | |

## 4.2. RESULTS

### 4.2.1. RELIABILITY AND VALIDITY:

We used the principal component analysis (PCA) and the confirmatory factor analysis (CFA) in order to evaluate the reliability of the construct.

- Principal Component Analysis (PCA)

The KMO (Kaiser-Meyer-Olkin) index measures proportion of variance among variables that might be common variance.

Using IBM SPSS Statistics, we have a value of 0,908 (Figure. 2) which represents a high level of sampling adequacy for factor Analysis.

Also, the Bartlett's Test of Sphericity (Figure.2) presents the validity and suitability of the responses collected to the problem being addressed through the survey. Considering a 95% level of significance ($\alpha = 0.05$), the p-value (Sig.) of .000 is under 0.05, and therefore the Factor Analysis is valid. We conclude that there may be statistically significant interrelationship between variables.

|  | Factor | | | |
|---|---|---|---|---|
|  | 1 | 2 | 3 | 4 |
| OA2 | 0,722 | 0,211 | 0,043 | 0,397 |
| OA4 | 0,697 | 0,218 | 0,261 | 0,346 |
| OA3 | 0,678 | 0,085 | 0,238 | 0,420 |
| OA6 | 0,654 | 0,346 | 0,414 | 0,016 |
| OA7 | 0,648 | 0,397 | 0,261 | 0,029 |
| OA1 | 0,640 | 0,162 | 0,044 | 0,222 |
| OA8 | 0,624 | 0,399 | 0,479 | 0,034 |
| OA5 | 0,618 | 0,330 | 0,226 | 0,154 |
| AC2 | 0,268 | 0,870 | 0,172 | 0,228 |
| AC3 | 0,313 | 0,862 | 0,151 | 0,205 |
| AC1 | 0,246 | 0,707 | 0,334 | 0,247 |
| AC4 | 0,553 | 0,598 | 0,213 | 0,098 |
| KMS2 | 0,169 | 0,140 | 0,875 | 0,101 |
| KMS1 | 0,156 | 0,205 | 0,739 | 0,290 |
| KMS3 | 0,252 | 0,340 | 0,604 | 0,219 |
| KMS4 | 0,403 | 0,095 | 0,523 | 0,397 |
| KMS7 | 0,247 | 0,279 | 0,015 | 0,789 |
| KMS6 | 0,247 | 0,154 | 0,334 | 0,721 |
| KMS5 | 0,167 | 0,149 | 0,431 | 0,707 |

Extraction method : Principal Axis factoring
Rotation method : Varimax with Kaiser normalisation.

Figure 2.  KMO and Bartlett's Test

Figure 3. Below shows that from the fourth factor on, the line is almost flat, meaning the each successive factor is accounting for smaller and smaller amounts of the total variance.

This is coherent with the number of factors in our proposed model.

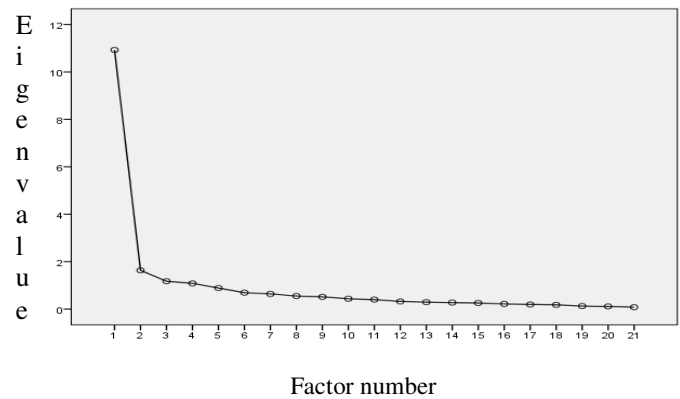

Figure 3. Scree Plot

Then we used the rotated matrix in order to identify the variables corresponding to each factor. In grey, we see the highest values of variables' correlations with factors. We can conclude that each factor is represented by these corresponding variables (Figure 4.)

| Kaiser-Meyer-Olkin Measure of Sampling Adequacy. | | ,908 |
|---|---|---|
| Bartlett's Test of Sphericity | Approx. Chi-Square | 1828,269 |
| | df | 171 |
| | Sig. | ,000 |

Figure 4. Rotated factor matrix

Based on this matrix, we can conclude the correspondence between the items of our framework and the four factors as below (Table 4.):

Table 4. Correspondence between the items of our proposed Framework and the four factors

| Item | Corresponding factor |
|---|---|
| Organizational agility | Factor 1 |
| Absorptive capacity | Factor 2 |
| Collaborative KMS | Factor 3 |
| Decision-oriented KMS | Factor 4 |

- ▪ Confirmatory Factor Analysis (CFA)

We use a confirmatory factor analysis (CFA) in order to confirm the factor structure extracted earlier in the exploratory analysis (PCA).

Specific metrics were adopted to determine goodness of fit. We compared the values calculated using IBM SPSS AMOS Software with the thresholds from Hu and Bentler (1999) [55]. Especially, the measure of RMSEA indicates that we should eliminate variables with the least correlations in order to represent well the corresponding latent factors.

Therefore, for an acceptable RMSEA = 0.093 which is below 0.10, factors 1, 2, 3 and 4 contains respectively the following variables (OA2,3,4,6,7), (AC1,2,3), (KMS1,2,3) and (KMS5,6,7). The other measures and their significations are presented in the table 5. Below.

Table 5.  Measures and their significations

| Measure | Definition | Value | Signification in comparison to threshold |
|---|---|---|---|
| RMSEA | A measure of goodness of fit for statistical models [56]. | 0,093 | Acceptable : below 0,10 |
| CFI | Examining the discrepancy between the data and the hypothesized model [57]. | 0,925 | Good : upper to 0,90 |
| TLI | Indicate how much better a model fits the data compared to a baseline model where all variables are uncorrelated [58]. | 0,906 | Good : upper to 0,90 |

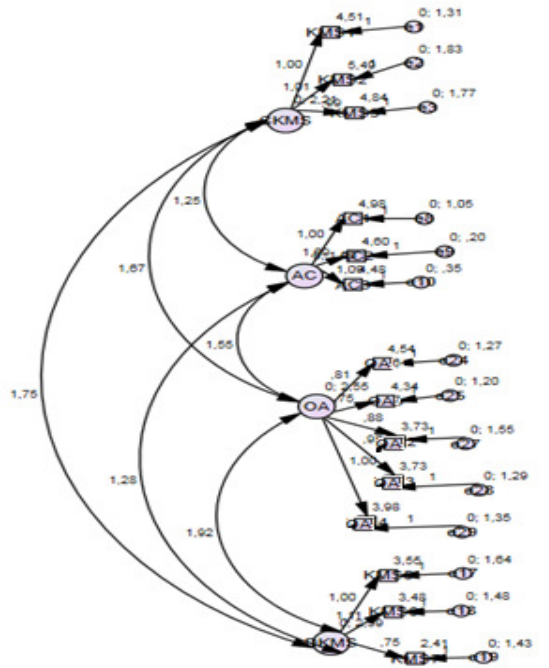

Figure 5. The factor confirmation model

Figure 5. shows the path diagram that represents the factor confirmation mode. The four latent variables are manifested by the corresponding observed variables as described earlier.

### 4.2.2. HYPOTHESIS TESTING: PLS

In order to verify the hypothesis of our model (H1, H2 and H3) between our latent variables, we use the Partial Least Squares method (PLS). This path modeling method, which was developed by Wold (1982), is a structural equation modeling (SEM) using a sequence of regressions in terms of weight vectors [59].

By using SmartPLS software, we calculated the PLS results on the model (Figure.6) with a maximum number of iterations set at 300 and a top criterion of (10^-7).

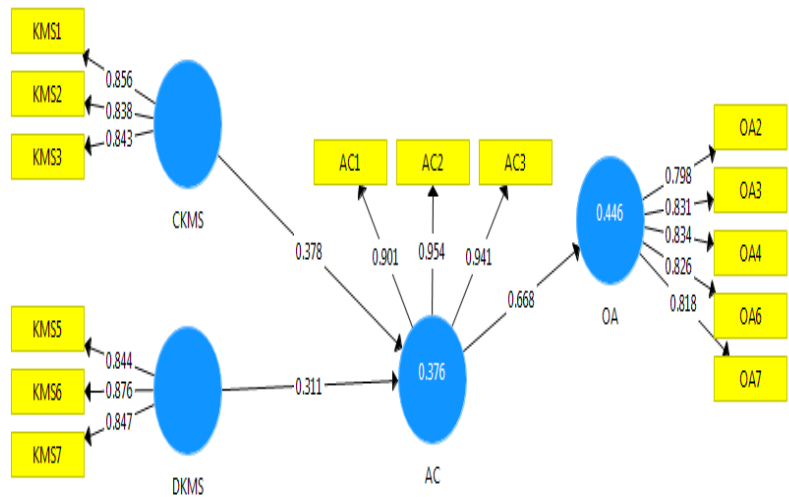

Figure 6. The PLS-SEM results on our model

We can deduct that on one side, CKMS and DKMS have respectively a positive direct effect (the inner model loading) of 0.378 and 0.311 on AC which supports the H1 and H2 hypothesis. Also, CKMS and DKMS have a less positive indirect effect of 0.253 and 0.207 on OA (Figure 7.).

On the other side, AC has a remarkable impact of 0.668 on OA which supports our third hypothesis H3.

Path Coefficients

|  | AC | CKMS | DKMS | OA |
|---|---|---|---|---|
| AC |  |  |  | 0,668 |
| CKMS | 0,378 |  |  |  |
| DKMS | 0,311 |  |  |  |
| OA |  |  |  |  |

Indirect Effects

|  | AC | CKMS | DKMS | OA |
|---|---|---|---|---|
| AC |  |  |  |  |
| CKMS |  |  |  | 0,253 |
| DKMS |  |  |  | 0,207 |
| OA |  |  |  |  |

Figure 7. Direct and indirect effects

# 5. CONCLUSIONS AND LIMITATIONS

This study has allowed exploring the relationship between the use of knowledge management systems and the improvement of firm's agility. Indeed, through the survey conducted, the results have supported the positive impact of knowledge management systems either directly on absorptive capacity or indirectly by enhancing the firm's organizational agility.

In conclusion, either the collaborative and decision-oriented knowledge management systems influence positively on improving firm's dynamic capabilities.

Otherwise, our sample size was the main limitation of our study due to the realized response rate. The goodness of our model fit would be better with a larger sample.

# 6. RESEARCH PERSPECTIVES

Future research will tackle specific areas by focusing on one chosen country, or narrowing the study on SMEs for instance in a particular sector.

Another perspective of our research would be to enlarge the study scope by adding new parameters and factors which may influence the absorptive capacity or the organizational agility of the firm.

Moreover, our proposed Framework may be extended by integrating the firm's sustainable performance as a final output.

## AUTHORS

M. Mohamed Amine Marhraoui is a PhD candidate at ENSIAS (National Higher School for Computer Science and System Analysis) Rabat Mohammed V University, Morocco. Within TIME Laboratory (Information Technology and Management), his thesis subject tackles the impact of Information Technology on firm's organizational agility and the mediating role of agility in enhancing enterprise's performance.

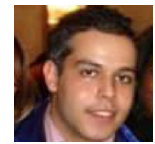

M. Marhraoui has graduated from ENSEIRB (Telecommunications, National School of electronics, computer and telecommunications of Bordeaux, France, 2007) and obtained an MSc in Commercial Engineering and Project Management from INSEEC Business School (Paris, France, 2011). M. Marhraoui has worked as a consultant in Information Technology management, organization and project management in different sectors (energy, industry and banking).

PR A. EL Manouar is Ph. D Economics; He has obtained diploma of a Bachelor in Economics and Law from University Mohammed V (Rabat, Morocco, 1980), a Master of Economics (Money, Finance and Banking, University of Clermont-Ferrand, France, 1981), a M. Sc. Economics (Money and Macroeconomic policy, Department of Economics, University of Montréal, Canada 1983) and a Ph. D Economics (Macroeconomic and international finance, Department of Economics, University of Montréal, Canada, 1990).He is now a Professor at ENSIAS (University Mohammed V, Rabat, Morocco) where he is teaching courses in economics, finance and electronic management. He accumulated a long teaching experience (almost 30 years) in different institutions (Morocco, Canada and France). PR A. EL Manouar's research interest covers new economy and ITC economics (Information Economy, digital economy and E-business) and Classical economics (Industrial economy, Macroeconomics and Monetary Theory, Corporate Finance, Economic Relations and International Finance). 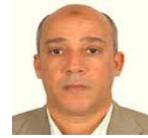